\documentclass[aps,prl,twocolumn,superscriptaddress,showpacs]{revtex4}

\usepackage{graphicx}

\bibliography{100}

\begin{document}

\title{\textbf{Resonant Plasmon--Soliton Interaction}}

\author{Konstantin Y. Bliokh}
\affiliation{Nonlinear Physics Centre, Research School of Physical Sciences and
Engineering, Australian National University, Canberra ACT 0200, Australia}

\author{Yury P. Bliokh}
\affiliation{Physics Department, Technion-Israel Institute of Technology, Haifa 32000,
Israel}

\author{Albert Ferrando}
\affiliation{Nonlinear Physics Centre, Research School of Physical Sciences and
Engineering, Australian National University, Canberra ACT 0200, Australia}
\affiliation{Interdisciplinary Modeling Group, InterTech. Departament d'\`{O}ptica,
Universitat de Val\`{e}ncia. Dr. Moliner, 50. E-46100 Burjassot (Val\`{e}ncia), Spain}
\affiliation{IUMPA, Universitat Polit\`{e}cnica de Val\`{e}ncia, E-46022 Val\`{e}ncia,
Spain.}

\begin{abstract}

We describe an effective resonant interaction between two localized wave modes of
different nature: a plasmon-polariton at a metal surface and a self-focusing beam
(spatial soliton) in a non-linear dielectric medium. Propagating in the same direction,
they represent an exotic coupled-waveguide system, where the resonant interaction is
controlled by the soliton amplitude. This non-linear system manifests hybridized
plasmon-soliton eigenmodes, mutual conversion, and non-adiabatic switching, which offer
exciting opportunities for manipulation of plasmons via spatial solitons.

\end{abstract}

\pacs{73.20.Mf, 42.65.Jx, 42.79.Gn}

\maketitle

\paragraph{Introduction.---} Plasmonics is an important, quickly developing area of modern physics
which offers promising applications in nano-optics and electronics [1--4]. It deals with the so-called surface plasmon-polaritons [5], i.e. collective
oscillations of the electromagnetic field and electrons which propagate along a
metal-dielectric surface and decay exponentially away from the surface.

Plasmons are characterized by frequency $\omega$ and propagation constant along the
interface, $k_p = k_p (\omega) > k$ ($k=\sqrt{\varepsilon_0}\omega/c$ is the wave number
in the dielectric medium with the permittivity $\varepsilon_0$). Since $k_p>k$, plasmons
can only interact resonantly with evanescent electromagnetic waves in the dielectric
medium [1--4]. Accordingly, there are two main methods for excitations of plasmons: (i)
via the evanescent wave generated at the total internal reflection [6]; and (ii) via a
periodic structure producing evanescent modes [4,7]. The first method can be modified if
there is a dielectric waveguiding layer parallel to the metal-dielectric interface.
Propagating modes of the waveguide have evanescent tails outside the waveguide. They can
interact resonantly with plasmons at the metal surface, that is used for diagnosis of
layered dielectric films [8]. The plasmon channel can also be regarded as an effective
waveguide, so that this configuration can be assimilated to the problem of two coupled
waveguides.

In this Letter, we propose a new way of resonant interaction of plasmons with
electromagnetic waves. Instead of exploiting various inhomogeneities to generate
evanescent modes, we introduce a \textit{non-linear} dielectric medium. A nonlinear
dielectric admits self-focusing solutions (spatial solitons) [9] which may propagate
parallel to the metal-dielectric interface and, similarly to the modes of the dielectric
waveguide, also have evanescent tails and propagation constant $k_s > k$ [10]. A
`quasi-waveguide' is formed by the soliton profile which modifies the effective
dielectric constant of the non-linear medium. Hence, the soliton-plasmon configuration
couples effective linear and non-linear waveguides, as shown in Fig. 1, and an
interaction between the plasmon and soliton may occur at certain resonant parameters. In
contrast to previous studies in non-linear plasmonics, where effects of nonlinearity on
plasmon modes have been considered (see, e.g., [11]), here we examine interaction between
two spatially-separated modes: soliton and \emph{linear} plasmon. This system is
essentially self-influencing -- the coupling is controlled by the soliton
\textit{amplitude} rather than by the configuration parameters.

\begin{figure}[t]
\centering \scalebox{0.32}{\includegraphics{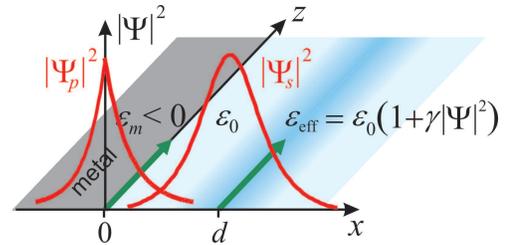}} \caption{(Color online.)
Plasmon-soliton system as two coupled waveguides.} \label{Fig2}
\end{figure}

\paragraph{Model.---} To create a model describing the plasmon-soliton system, Fig. 1,
we adopt several simplifying assumptions. First, the waves propagate along the $z$
coordinate and the wave electric field lies in the $(x,z)$ plane, so that the $y$
coordinate can be eliminated from further consideration. Second, we assume that the
dielectric non-linearity is localized around a certain distance $x=d$ from the metal
surface $x=0$, so that it practically does not affect the plasmon field. The distance $d$
is assumed to be larger than the characteristic widths of the plasmon and soliton fields,
which guarantees an exponentially-small overlapping of their tails. Hence, the coupling
is weak and can be treated perturbatively.

Uncoupled plasmon and soliton fields, $\Psi_p$ and $\Psi_s$ [12], can be represented as
$\Psi_p (x,z)=c_p (z) \psi_p (x)$ and $\Psi_s (x,z)=c_s (z) \psi_s (x, |c_s|)$, where
$c_{p,s}$ are the $z$-dependent amplitudes and $\psi_{p,s}$ are the transverse profiles
of the fields with the normalizations $\psi_p(0)=1$ and $\psi_s(d)=1$. The transverse
profile of the plasmon, $\psi_p$, represents two exponents decaying away from the
metal-dielectric interface [1,2] ($\psi_p=\exp(-\kappa_p x)$, $\kappa_p
=\sqrt{k_p^2-k^2}$ in the dielectric, at $x>0$), whereas the soliton profile is given by
[10] $\psi_s = {\rm sech}\left[\kappa_s (x-d)\right]$, $\kappa_s =k\sqrt{\gamma/2}|c_s|$,
Fig. 1. Here $\gamma$ is the parameter of nonlinearity of the medium [13]. We emphasize
that the transverse profile of the soliton depends on its amplitude $|c_s|$. In what
follows we assume that the amplitude varies slowly enough, so that one can use
quasi-stationary adiabatic approximation considering $|c_s|$ as a local parameter of the
problem.

Thus, the total wave field in the problem, $\Psi$, can be represented via the \textit{ansatz}
\begin{equation}\label{eq1}
\Psi(x,z) = c_p (z) \psi_p (x) + c_s (z) \psi_s (x,|c_s|)~.
\end{equation}
In the zero approximation, when the plasmon-soliton coupling is negligible, amplitudes
$c_p$ and $c_s$ are independent and obey, respectively, linear and non-linear oscillator
equations. In the first order approximation, the plasmon and soliton fields become
linearly coupled due to the spatial overlapping of $\Psi_p$ and $\Psi_s$ and the
transverse inhomogeneity of the medium along $x$ (i.e., the contrast of dielectric
indices at the metal-dielectric interface, $\varepsilon_m \neq \varepsilon_0$). This
results in the coupled oscillator equations for the amplitudes $c_p$ and $c_s$:
\begin{equation}\label{eq2}
c_p^{\prime\prime} + \beta_p^2 c_p = q(|c_s|) c_s~,~~c_s^{\prime\prime} + \beta_s^2
(|c_s|) c_s = q(|c_s|) c_p~.
\end{equation}
Here the prime stands for the derivative with respect to the dimensionless coordinate
$\zeta = kz$, $\beta_p = k_p/k > 1$ and $\beta_s =k_s/k \simeq 1 + \gamma |c_s|^2 /4$ (we
assume that $\gamma |c_s|^2\ll 1$) are the plasmon and soliton dimensionless propagation
constants, and $q\ll 1$ is the coupling coefficient.

To determine the coupling coefficient $q$, note that the right-hand side of the first
Eq.~(2), i.e. $q(|c_s|) c_s$, represents an external source that excites plasmons at the
metal surface. Hence, up to a numerical factor, it should be equal to the soliton field
at the metal surface, $\Psi_s|_{x=0}=c_s \psi_s |_{x=0}$. This yields the estimation
\begin{equation}\label{eq3}
q(|c_s|) \sim \psi_s|_{x=0} \simeq \exp\left(-k\sqrt{\gamma/2}|c_s|d\right)~,
\end{equation}
which is adopted below. Note that the weak-coupling approximation fails at small soliton
amplitudes.

Equations (1)--(3) represent the model describing the plasmon-soliton interaction. These
equations are reminiscent of the system of two weakly coupled nonlinear waveguides [14].
However, there are two essential peculiarities in the plasmon-soliton equations: (i) only
one subsystem is nonlinear; (ii) the coupling coefficient depends on the soliton
amplitude.

\begin{figure}[t]
\centering \scalebox{0.40}{\includegraphics{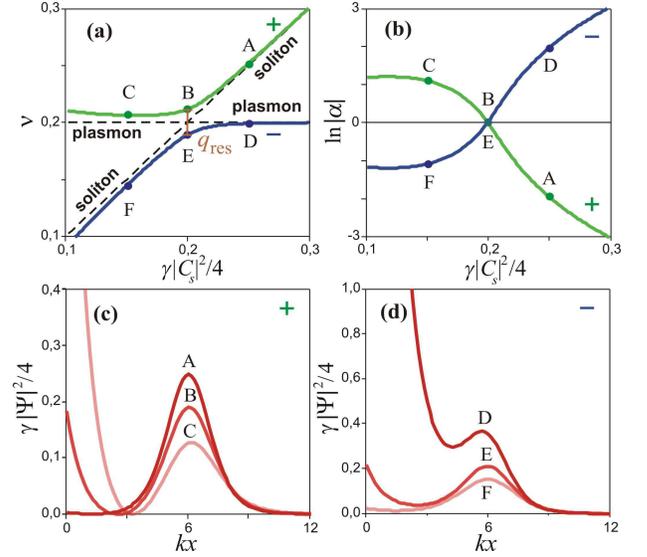}} \caption{(Color online.)
Collective modes in the plasmon-soliton system vs. the dimensionless soliton amplitude. (a) --
Eigenvalues $\nu^{\pm}$ Eq.~(7) (bold curves), eigenvalues of uncoupled modes $\nu_p$ and
$\nu_s$ (dashed lines). (b) -- Amplitude ratios $\alpha^{\pm}$ Eq.~(8). (c) and (d) --
Profiles of the total field $|\Psi(x)|^2$, Eq.~(1), corresponding to the points A--F at
curves in (a) and (b). The plasmon tails at $x<0$ are not depicted, since they are
determined by the properties of a particular metal. Parameters are $\nu_p=0.2$ and $kd=6$
($q_{\rm res}\simeq 0.02$).} \label{Fig2}
\end{figure}

An effective plasmon-soliton interaction and energy exchange occurs only near the
resonance $\beta_p = \beta_s$ which is achieved at the soliton amplitude $|c_s|_{\rm
res}=\sqrt{4(\beta_p - 1)/\gamma}$. Solitons with significantly larger or smaller
amplitude are uncoupled from the plasmon. In the vicinity of resonance, $|\beta_p -
\beta_s|\ll\beta_p$, Eqs. (2) can be simplified. Making substitution
$c_{p,s}(\zeta)=C_{p,s}(\zeta)\exp(i\zeta)$ and assuming that new amplitudes $C_{p,s}$
vary slowly ($|C_{p,s}^{\prime}|\ll |C_{p,s}|$ and $C_{p,s}^{\prime\prime}$ can be
neglected), we arrive at the following equations:
\begin{eqnarray}\label{eq4}
-i \left( \begin{array}{c} C_p \\ C_s \end{array} \right)^{\prime}= \left(
\begin{array}{cc} \nu_p & -q(|C_s|)/2 \\ -q(|C_s|)/2 & \nu_s(|C_s|) \end{array} \right)
\left( \begin{array}{c} C_p \\ C_s \end{array} \right).
\end{eqnarray}
Here
\begin{equation}\label{eq5}
\nu_p \equiv \beta_p - 1 \ll 1~,~~\nu_s \equiv \beta_s - 1 = {\gamma |C_s|^2}/{4} \ll 1
\end{equation}
are the small deviations of the dimensionless propagation constants from unity. The first
inequality (5) implies plasmons in the long-wave region of their spectrum close to the
light cone, whereas the second inequality indicates the weakness of nonlinearity .

Equation (4) has the form of a vector non-diffractive non-linear Schr\"{o}dinger equation
with a non-diagonal Hamiltonian typical in quantum two-level systems [15]. It possesses
the integral of motion $|C_p|^2+|C_s|^2 ={\rm const}$, which is associated with the
conservation of the total energy under the plasmon-soliton interaction.

\paragraph{Eigenmodes.---} Modes of Eq. (4) are obtained via
substitution $C_{p,s}(\zeta)=A_{p,s}\exp(i\nu\zeta)$. This yields the characteristic
equation determining the eigenvalues,
\begin{equation}\label{eq6}
\nu^{\pm}(|C_s|) = \frac{\nu_p+\nu_s(|C_s|) \pm
\sqrt{\left[\nu_p-\nu_s(|C_s|)\right]^2+q^2 (|C_s|)}}{2},
\end{equation}
and the ratios of the plasmon and soliton amplitudes determining the eigenvectors,
\begin{equation}\label{eq7}
\alpha^{\pm}(|C_s|)\equiv \frac{A_p}{A_s}=-\frac{q(|C_s|)}{2[\nu^{\pm}(|C_s|)-\nu_p]}~.
\end{equation}
Unlike linear systems, here the eigenvectors cannot be normalized arbitrarily because the
right-hand side of Eq. (7) depends on the amplitude $|C_s|=|A_s|$.

Equations (6) and (7) describe the collective (hybridized) plasmon-soliton modes
appearing due to the coupling. They have a standard form typical for a linear two-level
problem with close eigenvalues and small coupling [15]. In our case, the problem is
essentially non-linear and the properties of the solution depend on the soliton
\textit{amplitude} $|C_s|=|A_s|$. In other words, the soliton amplitude plays the role of
the \textit{driving parameter}.

Figure 2 shows the eigenvalues $\nu^{\pm}$, Eq.~(6), together with $\nu_p$ and $\nu_s$,
amplitude ratios $\alpha^{\pm}$, Eq.~(7), and field profiles $|\Psi(x)|^2$, Eq.~(1), of
the collective plasmon-soliton modes as dependent on the soliton amplitude $|C_s|$. In
the interaction region, $|\nu_p-\nu_s|\lesssim q_{\rm res}$, the eigenvalues avoid
crossing (the minimal frequency gap $q_{\rm res}\equiv q(|C_s|_{\rm
res})=\exp\left(-kd\sqrt{2\nu_p}\right)$ is achieved at resonance), and eigenmodes are
collective excitations with $|A_p|\sim |A_s|$. At resonance, $\nu_s = \nu_p$, the two
modes show their closest intensity distributions (B and E) corresponding to the symmetric
and antisymmetric combination of plasmon and soliton states with equal amplitudes. Away
from the interaction region, $|\nu_p-\nu_s|\gg q_{\rm res}$, the eigenvalues $\nu^{\pm}$
tend to the asymptotes $\nu_p$ and $\nu_s$ representing pure plasmon and soliton states.
Accordingly, the energy is concentrated basically in either the plasmon (D, C) or the
soliton (A, F) channel. When passing through the interaction region along the same branch
in Fig. 1a, the eigemodes exchange the amplitudes, Fig.~2b, and a near-soliton mode
transmutes into the plasmon one and vice versa. This offers various possibilities for
energy conversion between the plasmon and soliton channels.

\paragraph{Dynamics and dissipation.---} First, we consider the case when the soliton
channel is excited at the input with a near-resonance amplitude $C_s(0)$:
\begin{equation}\label{eq8}
C_p(0)=0~,~~\left|C_s(0)-|C_s|_{\rm res}\right| \ll |C_s|_{\rm res}~.
\end{equation}
Pure soliton is not an eigenmode, and initial conditions (8) excite a mixture of ``$+$''
and ``$-$'' eigenmodes with close frequencies. In a linear problem this would lead to a
superposition of two modes and a harmonic low-frequency beating. But in the
plasmon-soliton system, the superposition principle is invalid, and the solution of Eq.
(4) with Eq. (8) represent a self-modulated non-linear beating  causing a periodic mutual
conversion of energy between the soliton and plasmon channels, Fig.~3. The maximal
conversion amplitude is reached at a certain value $C_{s}(0)_{\rm max} > |C_s|_{\rm
res}$. It corresponds to the maximal beating period $\Delta\zeta\sim 2\pi/q_{\rm res}$
and it drops abruptly for higher $C_s(0)$. The beating amplitude decreases with the
resonant coupling $q_{\rm res}$ decrease, Fig.~3a and b.

\begin{figure}[t]
\centering \scalebox{0.41}{\includegraphics{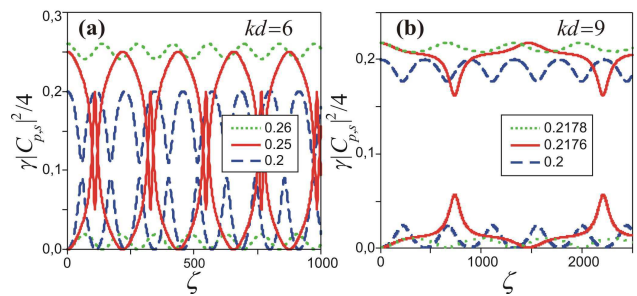}} \caption{(Color online.)
Nonlinear beating of the plasmon (lower curves) and soliton (upper curves) dimensionless amplitudes
after the excitation of the soliton channel near resonance. Eqs. (4) and (8) are
solved numerically with $\nu_p=0.2$ and values of $\gamma |C_s(0)|^2/4$ as indicated in
the boxes. Values of $kd$ correspond to $q_{\rm res}\simeq 0.02$ (a) and $q_{\rm
res}\simeq 0.003$ (b).} \label{Fig3}
\end{figure}

Second, soliton-plasmon conversion occurs under manipulations with the soliton amplitude
-- the driving parameter of the system. One could expect that slow changes of the soliton
amplitude will result in an adiabatic transformations of the eigenmodes along the ``$+$''
or ``$-$'' dispersion curves, Fig.~2a, and the soliton will metamorphose into plasmon
when passing over the resonance. For instance, the soliton is excited with an
off-resonance amplitude corresponding to the zone A in Fig.~2, i.e.
\begin{equation}\label{eq9}
C_p(0)=0~,~~C_s(0)-|C_s|_{\rm res} \sim |C_s|_{\rm res}~.
\end{equation}
These initial conditions approximately match the ``$+$'' eigenmode. Then, we introduce a
small absorption in the medium which makes the soliton amplitude slowly decreasing along
$\zeta$, $|C_s|=|C_s|(\zeta)$. The losses in the plasmon and soliton channels are
described by adding small imaginary parts $\sigma_{p,s}$ to their propagation constants:
\begin{equation}\label{eq10}
\nu_{p,s}\rightarrow \tilde{\nu}_{p,s}=\nu_{p,s} + i\sigma_{p,s}~,~~|\sigma_{p,s}|\ll
q_{\rm res}~.
\end{equation}
The weak dissipation (10) does not affect the local characteristics of the waves, which
change adiabatically with $|C_s|(\zeta)$ according to the eigenmodes (6) and (7).

\begin{figure}[tbh]
\centering \scalebox{0.43}{\includegraphics{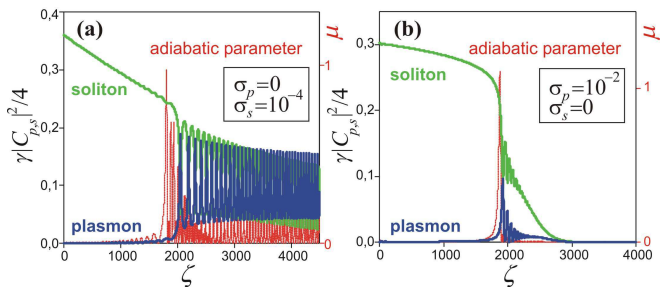}} \caption{(Color online.)
Adiabatic dynamics in the plasmon-soliton system with small dissipation, Eqs. (4),
(9)--(11), $\nu_p=0.2$ and $kd=6$. Initial soliton (9) attenuates slowly until it reaches
the resonant amplitude. Then, only a part of the energy is abruptly transferred to the
plasmon channel ($\mu_{\rm res} \gtrsim 1$) with a subsequent beating in the mix of
``$+$'' and  ``$-$'' eigenmodes, cf. Fig.~3. The value of the adiabatic parameter
$\mu(\zeta)$ (thin dotted line) indicates the effectiveness of the energy exchange between the eigenmodes.}
\label{Fig4}
\end{figure}

With Eq.~(10), this leads to the transformation of states as $A\rightarrow B\rightarrow
C$, Fig. 2, i.e. almost 100\% conversion of energy from the soliton to the plasmon
channel. This would be a perfect switch, but the non-adiabatic Landau-Zener transitions
between the ``$+$'' and ``$-$'' eigenmodes may appear in the vicinity of resonance,
$|\nu_p-\nu_s|\lesssim q_{\rm res}$ [15,16]. The probability of the Landau- Zener
transitions can be estimated using the parameter of adiabaticity $\mu$ [17]:
\begin{equation}\label{eq11}
\mu = {|\nu^{\pm\prime}|}/{|\nu^+ -\nu^- |^2}~.
\end{equation}
If $\mu\ll 1$, the ``$+$'' and ``$-$'' eigenmodes evolve independently according to the
adiabatic evolution. On the contrary, if $\mu \gg 1$, the wave undergoes diabatic
evolution and practically all the energy stored in the ``$+$'' eigenmode is converted
into the ``$-$'' one when passing the resonance. This means the transformation of states
as $A\rightarrow B+E\rightarrow F$, Fig.~2, i.e. the non-adiabatic transition keeps the
energy in the soliton channel. In the intermediate case, $\mu \sim 1$, the output wave
will present a mixture of the ``$+$'' and ``$-$'' modes: $A\rightarrow C+F$.

Surprisingly, the adiabatic parameter (11) cannot be made small in the plasmon-soliton
system, no matter how weak the dissipation is. The point is that the soliton amplitude
$|C_s|$ dramatically changes in the vicinity of resonance due to soliton-to-plasmon
transmutation along the eigenmode, and the adiabatic regime is never achieved. This is a
purely nonlinear effect -- the driving parameter is a part of the solution. Numerical
simulations of Eq.~(4) with Eqs.~(9)--(11) confirm this conclusion, Fig.~4. In the
vicinity of resonance, $\mu_{\rm res} \gtrsim 1$ which results in a partial
soliton-to-plasmon switching and the Landau-Zener transition with the consequent
nonlinear beating between ``$+$'' and ``$-$'' eigemodes. Smaller values of the coupling
$q_{\rm res}$ result in higher $\mu_{\rm res}$ and smaller part of energy transferred to
the plasmon.

\paragraph{Discussion.---} To summarize, we have shown the possibility of resonant interaction
between a plasmon-polariton at a metal surface and a parallel self-focusing beam in a
nonlinear dielectric. A simple two-level model reveals hybridized plasmon-soliton
eigenmodes and their complex non-linear dynamics which offers plasmon excitation and
control using spatial solitons.

An effective soliton-to-plasmon coupling and energy conversion can be achieved in the far
IR frequency range (the wavelength in vacuum is $\lambda_0 \sim 5\div 10{\rm \mu m}$), by
using a gold interface with a non-linear dielectric (e.g., a chalcogenide glass,
$\varepsilon_0 \sim 5 \div 10$). In this range, plasmons are characterized by the
propagation constant close to unity, $\nu_p \sim 10^{-3}$ and relatively small
dissipation $\sigma_p \sim 0.1$. Solitons with $\nu_s=\gamma |c_s|^2/4 \sim 10^{-3}$ can
be produced with typical Kerr nonlinearities. Transverse evanescent tails of plasmon and
soliton are determined by $\kappa_p \sim \kappa_s \sim 0.1 {\rm \mu m}^{-1}$, and
propagation of the soliton at the distance $d\sim 20\div 40 {\rm \mu m}$ ($kd \sim 50\div
100$) from the metal surface yields the coupling coefficient $q_{\rm res}\sim 10^{-1}\div
10^{-2}$. Although the plasmon dissipation is large enough, $\sigma_p \gtrsim q_{\rm
res}$ [18], we have found, by numerical simulations of Eqs.~(4) with (10), that our
conclusions on the plasmon-soliton coupling and dynamics (e.g., Fig.~4b) remain
qualitatively true at these realistic parameters.

This work was supported by contracts  FIS2005-01189 and TIN2006-12890 from the Government
of Spain, and by the Australian Research Council.


\begin{references}
\bibitem{12}{By the field $\Psi$ we imply the $x$-component of the wave electric field or the $y$-component
of the magnetic field.}
\bibitem{13}{We introduce the nonlinearity $\gamma$ so that the effective dielectric constant
of the medium is $\varepsilon_{\rm eff} = \varepsilon_0(1 + \gamma |\Psi|^2)$.}
\end{references}
\end{document}